\documentclass[prd,
superscriptaddress,showpacs,nofootinbib,%
tightenlines ]{revtex4}
\usepackage{epsfig}
\usepackage{amssymb}

\newcommand{\ben}{\begin{displaymath}}
\newcommand{\een}{\end{displaymath}}
\newcommand{\be}{\begin{equation}}
\newcommand{\ee}{\end{equation}}
\newcommand{\bea}{\begin{eqnarray}}
\newcommand{\eea}{\end{eqnarray}}
\begin{document}
\title{Asymptotic freedom in massive Yang-Mills theory}
\author{J.~Gegelia}
\affiliation{Institut f\"ur Kernphysik, Johannes
Gutenberg-Universit\"at, J.J. Becherweg 45, \\ D-55099 Mainz,
Germany \\ and \\ High Energy Physics Institute of TSU, Tbilisi,
Georgia}

\date{\today}

\begin{abstract}
An effective field theory model of the massive Yang-Mills theory
is considered. Assuming that the renormalized coupling constants
of 'non-renormalizable' interactions are suppressed by a large
scale parameter it is shown that in analogy to the non-abelian
gauge invariant theory the dimensionless coupling constant
vanishes logarithmically for large values of the renormalization
scale parameter.

\end{abstract}





\maketitle

The standard model (SM) is an established consistent theory of
strong, electromagnetic and weak interactions. It describes most
of the known phenomena in elementary particle physics.  The modern
point of view is to think of the SM as an effective field theory,
``low-energy approximation to a deeper theory that may not even be
a field theory, but something different like a string theory''
\cite{Weinberg:mt}. While the effective Lagrangian consists of an
infinite number of terms, the non-renormalizable interactions are
suppressed by powers of a large scale. This makes the
contributions of these interactions negligible for energies much
lower than the large scale. Renormalizability in the traditional
sense is not considered as a fundamental principle but rather a
property of the leading order approximation to the full effective
theory.

The SM takes gauge invariance as a starting point.  In modern
string theories \cite{Green:mn} one first notices a state of mass
zero and unit spin among the normal modes of a string, and then
from that deduces the gauge invariance of the effective field
theory that describes such particles \cite{Weinberg:mt}. In modern
approach one usually takes gauge invariance as the starting point.
Referring again to Weinberg's book, ``It is too soon to tell which
of these two alternatives corresponds to the logical order of
nature'' \cite{Weinberg:mt}. Taking into account the above
considerations it is not evident that the gauge invariance
principle should be applied to weak interaction as it is not
mediated by massless vectors. As the fundamental nature of both
the gauge invariance and the renormalizability seems to be
questionable, the role of the scalar sector of the electro-weak
theory could be also questioned, especially as the experimental
status of the scalar Higgs particle remains still unclear.
Therefore it seems interesting to have a closer look to effective
field theory models of massive vector mesons which are not based
on the concept of gauge invariance.

In the present work an effective field theory model of the massive
vector mesons interacting with fermions is considered. I do not
deal here with the most general effective Lagrangian, but instead
add the mass term of the vector mesons to the standard QCD
Lagrangian \cite{Yndurain:1999ui}. Next I add an infinite number
of non-renormalizable interactions which are required to absorb
the divergences generated by loop diagrams of the perturbation
theory. I assume that all coupling constants of non-renormalizable
interactions are suppressed by powers of a Large scale parameter
$\Lambda$. Under this assumption I study the renormalization group
behavior of the dimensionless coupling constant to one-loop order.
The aim of this work is to investigate if the asymptotic freedom
of the non-abelian gauge theories
\cite{Politzer:1973fx,Gross:1973ju} persist when the explicit mass
term is introduced in the Lagrangian  (For earlier works on
asymptotically free massive Yang-Mills models see
Refs.~\cite{Hsu:1974ix,Bardeen:1978cz}). Notations below closely
follow the book by Yndur\'ain \cite{Yndurain:1999ui}.

Let us consider an EFT model described by the following bare
Lagrangian:
\begin{equation}
{\cal L} ={\cal L_{QCD}}+\frac{M^2}{2}\,\sum_b B^\mu_a B_{a
\mu}+{\cal L}_1\,, \label{lagrangianQCD0}
\end{equation}
where
\begin{equation}
{\cal L_{QCD}} =\sum_q \sum_j \bar q^j \left( i \partial
\hspace{-.6em}/\hspace{.1em}-m_q\right) q^j+g\,\sum_q\sum_{i k
a}\bar q^i \gamma_\mu t^a_{ik}q^k B_a^\mu -\frac{1}{4}\,\sum_a\,
G^{\mu\nu}_a G_{a\mu\nu}\,, \label{lagrangianQCD}
\end{equation}
is the standard QCD Lagrangian \cite{Yndurain:1999ui} and ${\cal
L}_1$ contains an infinite number of terms of non-renormalizable
interactions which are necessary for the cancelation of
divergences of loop diagrams.

Let us investigate the renormalization group behavior of
renormalized coupling constant $g_R$ by analyzing the
$B\bar\psi\psi$ Green's function using dimensional regularization
(with parameter $n$) in combination with the MS scheme.

To carry out the renormalization let us introduce the renormalized
quantities
\begin{equation}
q^j=Z_q^{1/2}q^j_R\,, \ B^\mu_{a}=Z_B^{1/2}B^\mu_{a,R}\,, \ m_q
=Z_m^{-1}\, m_{q,R}\,, \ M^2=Z_M^{-1}\, M_{R}^2\,, \ g=Z_g^{-1}g_R
\,,\cdots\,, \label{renquantities}
\end{equation}
(where $\cdots$ stands for parameters hidden in ${\cal L}_1$) and
choose $Z_q$, $Z_B$, $Z_m$, $Z_M$, $Z_g$, $\cdots$ such that all
Green's functions are finite order-by-order in loop expansion.

The sum of one-particle irreducible diagrams contributing in
vector-meson propagator is parameterized as $-i\,\Pi^{\mu\nu}$ and
to one loop order consists of a fermion and vector meson
contributions.

Divergent part of the contribution of all $n_f$ fermions reads:
\begin{equation}
\Pi^{\mu\nu}_{F;ab}=-2\,T_F\delta_{ab}
\,\frac{g_R^2}{16\,\pi^2}\,\left( -g^{\mu\nu}q^2+q^\mu
q^\nu\right)\,\frac{2}{3}\,N_\epsilon \,n_f\,,\label{gseFdivp}
\end{equation}
where $N_\epsilon =2/(4-n)$ and $T_F=1/2$. This has to be canceled
by counter-term contribution generated by $Z_B$.

Divergent part of the contribution of vector fields is given as
\begin{equation}
\Pi^{\mu\nu}_{ab;g}=\frac{C_A\,\delta_{ab}\,g_R^2}{32\,\pi^2}
\,N_\epsilon\, \left[ -9\, M^2 g^{\mu \nu}+7 \left( q^{\mu }
q^{\nu }-g^{\mu \nu } q^2\right) -\frac{\left( q^{\mu } q^{\nu
}-g^{\mu \nu } q^2\right) \left(
14\,M^2\,q^2+q^4\right)}{12\,M^4}\right]\,. \label{vsevcalc}
\end{equation}
The first term in square brackets in Eq.~(\ref{vsevcalc}) is
canceled by mass counter-term, second is canceled by $Z_B$ and the
third term is taken care by counter-term contributions generated
by higher-order terms hidden in ${\cal L}_1$.

Field redefinition constant obtained from Eqs.~(\ref{gseFdivp})
and (\ref{vsevcalc}) reads
\begin{equation}
Z_B=1+\frac{g_R^2}{16\,\pi^2}\,N_\epsilon\,\left[
\frac{7\,C_A}{2}-\frac{4\,T_F\,n_f}{3}\right]\,. \label{Brc}
\end{equation}

The sum of one-particle irreducible diagrams contributing in
fermion propagator
is given by
\begin{equation}
\Sigma_{ji}^{div}
(p)=\frac{g_R^2}{16\,\pi^2}\,\mu^{4-n}\,C_F\,\delta_{ji}\,N_\epsilon\,\biggl[
3\,m-\frac{m^3}{M^2}+\frac{3\,m^2}{2\,M^2}\,p
\hspace{-.45em}/\hspace{.1em}-\frac{p^2}{2\,M^2}\,p
\hspace{-.45em}/\hspace{.1em}\biggr]\,. \label{fsediv}
\end{equation}
The first two terms in square brackets in Eq.~(\ref{fsediv}) are
canceled by mass counter-term, third is canceled by $Z_q$ and the
fourth term is taken care by counter-term contributions generated
by higher-order terms hidden in ${\cal L}_1$.

Field redefinition constant obtained from Eq.~(\ref{fsediv}) reads
\begin{equation}
Z_q=1+\frac{3\,m^2\,g_R^2}{32\,\pi^2\,M^2}\,C_F\,N_\epsilon\,.
\label{frc}
\end{equation}
This gives the corresponding counter-term diagram contribution in
$B\bar\psi\psi$ vertex function
\begin{equation}
i\,\Gamma^{(1) \mu, Z_q}_{ija} =
\frac{3\,i\,g_R^3\,m^2}{32\,\pi^2\,M^2}\,C_F\,t_{ij}^{a}\,N_\epsilon\,\mu^{4-n}\,\gamma^\mu
\,. \label{Zqctd}
\end{equation}

Two one-loop diagrams contribute in $B\bar\psi\psi$ vertex
function. Keeping only momentum independent divergent
part\footnote{The momentum dependent divergent parts are canceled
by counter-term contributions generated by corresponding terms in
${\cal L}_1$.} contributing in the renormalization of $g$ the
first diagram gives
\begin{eqnarray}
i\,\Gamma^{(1) \mu}_{ija} &=&-\frac{3\,i}{64\,\pi^2}
\,\frac{m^2}{M^2}\,g_R^3\,C_A\,t^a_{ji}\,\mu^{4-n}\,N_\epsilon\,\gamma^\mu
\,. \label{gffbdiv}
\end{eqnarray}
Analogous divergent part of the second diagram reads
\begin{equation}
i\,\Gamma^{(2) \mu}_{ija}=-\frac{3\,
i}{32\,\pi^2}\,\frac{m^2}{M^2}\, g_R^3\,\left(
C_F-\frac{1}{2}\,C_A\right)\,t^a_{ji}\,\mu^{4-n}\,N_\epsilon\gamma^\mu\,.
\end{equation}

For the momentum-independent divergent part of the sum of two
diagrams we obtain
\begin{eqnarray}
i\,\Gamma^{(1) \mu}_{ija}+i\,\Gamma^{(2) \mu}_{ija}
&=&-\frac{3\,i\,g_R^3\,m^2}{32\,\pi^2\,M^2}\,C_F\,t_{ij}^{a}\,N_\epsilon\,\mu^{4-n}\,\gamma^\mu
\,. \label{gffbpcdiv}
\end{eqnarray}
The contributions of two loop diagrams, Eq.~(\ref{gffbpcdiv}), is
exactly canceled by counter-term diagram contribution,
Eq.~(\ref{Zqctd}), and the resulting value for $Z_g$ at one-loop
order is
\begin{equation}
Z_g=Z_B^{-1/2}=1- \frac{g_R^2}{32\,\pi^2}\,N_\epsilon\,\left[
\frac{7\,C_A}{2}-\frac{4\,T_F\,n_f}{3}\right]\,. \label{dg}
\end{equation}

For the corresponding running coupling constant $\alpha
=g_R^2/(4\,\pi)$ one obtains
\begin{equation}
\alpha(\mu)= \frac{\alpha\left( \mu_0\right)}{1+\frac{\alpha\left(
\mu_0\right)}{2\,\pi}\,\ln\frac{\mu}{\mu_0}\,\left[
\frac{7\,C_A}{2}-\frac{4\,T_F\,n_f}{3} \right]}\,.\label{rcc}
\end{equation}
The renormalized coupling $\alpha$ of  Eq.~(\ref{rcc}) vanishes
logarithmically for large $\mu$ if $n_f < 21\,C_A/4$.

To make sure that the considered model is self-consistent, I have
checked that the renormalization of the $BBB$ Green's functions
leads to the same results as given in Eqs.~(\ref{dg})-(\ref{rcc}).

To conclude, renormalization group behavior of the dimensionless
coupling constant of an effective field theory model of massive
Yang-Mills vector fields interacting with fermions has been
considered. This analysis has been performed under assumption that
the coupling constants of non-renormalizable interactions are
suppressed by powers of a large scale parameter $\Lambda$. The
renormalized coupling constant vanishes logarithmically for large
(but still $<<\Lambda$) values of the renormalization scale
parameter $\mu$ if the number of fermions $n_f < 21\,C_A/4$. This
is to be compared with $n_f < 11\,C_A/2$ in the gauge invariant
theory. For $\mu\sim\Lambda$ the contributions of the
non-renormalizable interactions can not be neglected, therefore
Eq.~(\ref{rcc}) can not be trusted for such values of $\mu$. The
same applies to QCD as well. According to modern understanding QCD
should be considered as the leading order approximation to an
effective field theory. For the energies where the contributions
of non-renormalizable interactions are no longer suppressed, the
asymptotic freedom of the strong interaction described by standard
renormalizable QCD Lagrangian requires further investigation. Of
course it is not evident that at such high energies the quantum
field theory approach is still meaningful.

\acknowledgments

The author would like to thank G.~Gabadadze for comments on the
manuscript. This work was supported by the Deutsche
Forschungsgemeinschaft (SFB 443).



\end{document}